\documentclass[a4paper,fleqn,usenatbib,useAMS]{mnras}
\usepackage{graphicx}   
\usepackage{amsmath}    
\usepackage{amssymb}    
\usepackage{lscape}
\newcommand{\vsini}[1]{$v\sin{i_s}$#1}
\newcommand{\sqvsini}[1]{$\sqrt{v\sin{i_s}}$#1}
\newcommand{\nobs}[1]{$N_{\mathrm{obs}}$#1}
\newcommand{\ms}[1]{m s$^{-1}$#1}
\def\ryan[#1]{\noindent{\textbf{\textcolor{red}{#1}}}}
\defcitealias{gillon16}{G16}
\defcitealias{gaudi07}{GW07}

\title[RM effect of Earth-like Planets]{Prospects for detecting the 
Rossiter-McLaughlin effect of Earth-like planets: the test case of 
TRAPPIST-1b and c}

\author[Cloutier \& Triaud]{Ryan Cloutier$^{1,2,3}$ 
\& Amaury H.~M.~J. Triaud$^{4}$ \\
$^{1}$Dept. of Astronomy \& Astrophysics, University of Toronto. \\
50 St. George Street, Toronto, Ontario, Canada,M5S 3H4\\
$^{2}$Centre for Planetary Sciences, Department of Physical \& Environmental Sciences, University of Toronto. \\ 
1265 Military Trail, Toronto, Ontario, Canada, M1C 1A4\\
$^{3}$Institut de recherche sur les exoplan\`{e}tes,D\'{e}partement de physique, Universit\'{e} de Montr\'{e}al. \\
2900 boul. Édouard-Montpetit, Montr\'{e}al Quebec, Canada, H3T 1J4\\
$^{4}$Institute of Astronomy, University of Cambridge. Madingley Road, Cambridge CB3 0HA, United Kingdom.\\}
\pubyear{2016}

\begin{document}
\label{firstpage}
\pagerange{\pageref{firstpage}--\pageref{lastpage}}
\maketitle

\begin{abstract}
The Rossiter-McLaughlin effect is the principal method of determining the sky-projected 
spin--orbit angle 
($\beta$) of transiting planets. Taking the example of the recently discovered TRAPPIST-1 
system, we explore how ultracool dwarfs facilitate the measurement of the spin--orbit angle 
for Earth-sized planets by creating an effect that can be an order of magnitude more ample 
than the Doppler reflex motion caused by the planet if the star is undergoing rapid rotation. 
In TRAPPIST-1's case we expect the Rossiter-McLaughlin semi-amplitudes to be $40-50$ \ms{} for the known 
transiting planets. 
Accounting for stellar jitter expected for ultracool dwarfs and instrumental noise, 
and assuming radial 
velocity precisions both demonstrated and anticipated for upcoming near-infrared spectrographs, we 
quantify the observational effort required to measure the planets' masses and spin--orbit angles. 
We conclude that if the planetary system is well-aligned then $\beta$ 
can be measured to a precision of $\lesssim 10^{\circ}$ if the spectrograph 
is stable at the level of 2 \ms{.} 
We also investigate the measure of $\Delta \beta$, the mutual inclination, 
when multiple transiting planets are present in the system. Lastly, we 
note that the rapid rotation rate of many late 
M-dwarfs will amplify the Rossiter-McLaughlin signal to the point where variations 
in the chromatic Rossiter-McLaughlin effect from atmospheric absorbers should be 
detectable.
\end{abstract}


\section{Introduction}
The Rossiter-McLaughlin effect is an anomalous radial velocity signal first observed 
in eclipsing binary systems and more recently 
in transiting exoplanetary systems \citep{rossiter24, mclaughlin24}. As opposing stellar 
limbs are systematically occulted by a transiting planet, the symmetry of stellar emission 
from each point on the stellar disc is broken as some starlight is blocked by the planet. 
If the star has any intrinsic rotation then the transit will cause an excess of 
particularly 
Doppler-shifted photons to be observed. This gives rise to the anomalous Rossiter-McLaughlin 
(RM) signal affecting radial velocity measurements made during transit. 
The resulting RM 
waveform is therefore sensitive to the planet's orbital trajectory across the stellar disc 
and hence to the alignment of the stellar spin 
axis with the normal to the planet's orbital plane. The RM effect is therefore suitable to 
measurement of the so-called sky-projected spin--orbit angle or simply spin--orbit 
angle as it will be referred to for the remainder of this paper \citep{ohta05, gimenez06}. 
Detection of the RM effect and hence the measurement of the 
spin--orbit angle is useful for probing the orbital history of a planetary system 
\citep{fabrycky09, triaud10, albrecht12} as well-aligned planets are predicted from 
models of terrestrial planet formation and highly misaligned 
systems may be reconciled with past dynamical events such as planet-planet interactions 
\citep{rasio96, weidenschilling96} or Kozai migration \citep{wu03}. 

Because the RM effect arises from the differential occultation of Doppler-shifted 
stellar limbs, the semi-amplitude of the RM effect 
scales with the projected stellar rotation velocity \vsini{}\footnote{We reserve the 
notation $i_s$ for the inclination of the stellar spin axis to the line-of-sight and $i_p$ 
for planetary orbital inclinations relative to the plane of the sky.} and is also dependent 
on the transit depth \citep{gaudi07}. In this way, giant planets around rapidly rotating 
stars become the optimal candidates for observing the RM effect. 
Following the scaling of the RM semi-amplitude with transit depth and \vsini{,} numerous 
studies of the RM effect for hot Jupiters have been performed 
\citep[e.g.][]{queloz00, triaud10, albrecht12, brown12}. 
Furthermore, the RM effect due to small 
rocky planets is often below the detection threshold of current velocimeters unless the 
planet's host star is also small. Fortunately, such planets are common around the 
M-dwarf class of small stars \citep{dressing13, dressing15a, gaidos16} 
and M-dwarf rotation periods appear to contain a significant population of fast rotators 
\citep[$P_{\mathrm{rot}} \lesssim 2$ days;][]{irwin11, mcquillan13a, mcquillan14}, especially 
among the closest M-dwarfs \cite{newton16a}, 
for which the RM effect of Earth-like planets may be sufficiently large to be detected.

In preparation for the ultracool dwarf transit survey instrument \emph{SPECULOOS} 
\citep{gillon13}, a pilot survey conducted by the \emph{TRAPPIST} telescope 
\citep{gillon11} was used to monitor the brightest ultracool dwarfs 
\citep[$T_{\mathrm{eff}}<2700$ K;][]{kirkpatrick95} in the Southern Hemisphere, 
like was done for Luhman-16AB \cite{gillon13b}. Recently 
\citep[][hereafter \citetalias{gillon16}]{gillon16} reported 
the discovery of a planetary system composed of three Earth-like planets, all 
transiting a nearby ultracool dwarf now known as TRAPPIST-1 (2MASS J23062928--0502285). 
This system represents a superlative opportunity 
to observe the RM effect of rocky planets. 
The planetary system contains a minimum of three small ($r_p < 1.2$ R$_{\oplus}$) transiting 
planets (transit depths $\gtrsim 0.67$\%) 
with orbital periods of $<20$ days, although the orbital period of the outermost 
planet is not unambiguously detected due to the small number of observed transits 
coupled with discontinuous observations. This makes the interpretation of the outermost 
planet indefinite. However, 
given the planets' small observed radii it is expected that the planets have bulk rocky 
compositions \citep{dressing15b, lopez14} although spectroscopic follow-up observations 
are likely required to characterize the planet masses thus constraining their bulk 
densities. The star TRAPPIST-1 is a late M-dwarf \citep[M7.5;][]{gizis00} with 
spectroscopically measured 
\vsini{}$=6 \pm 2$ km s$^{-1}$ \citep{reiners10b} (see Table~\ref{trappist1table} for 
a summary system properties). From this we estimate the RM 
semi-amplitudes to be $K_{\mathrm{RM}} \sim 40-50$ m s$^{-1}$ or an order-of-magnitude greater 
than the planets' expected Doppler semi-amplitudes\footnote{Sometimes referred to 
as the `orbital' semi-amplitude in the literature.}.

In this letter we investigate the radial velocity observations required to study 
the spin--orbit angles of Earth-like planets around cool stars using the TRAPPIST-1 
planetary system as our fiducial test case. In Sect.~\ref{methods} we explain our 
methods of constructing synthetic radial velocity timeseries, followed by an explanation 
in Sect.~\ref{model} of our model fitting procedures. Our results are then 
reported in Sect.\ref{results} followed by a discussion in Sect.~\ref{discussion} and 
conclusions in Sect.~\ref{conclusion}.

\section{Radial Velocity Timeseries Construction} \label{methods}
In order 
to investigate the effort required to recover the spin--orbit angles of the two innermost 
TRAPPIST-1 planets, we perform an extensive Monte-Carlo simulation of the 
expected radial velocity timeseries under realistic 
observing conditions, model the RM effect, and quantify the detection significance 
of the model parameters of interest including the spin--orbit angle. 
The details of our simulations are discussed below. 

\subsection{Window Function}
The window function defines the epochs at which radial velocity measurements are made. 
To measure the masses of the two innermost TRAPPIST-1 planets we construct window functions 
containing $N_{\mathrm{obs}}$ measurements made outside of transit so as not to be contaminated 
by the anomalous RM effect. The full width of the window functions span $\sim 6$ months 
assuming that dedicated observations of the star are being executed at most twice per night. 
The epochs from which 
we draw from are generated from the TRAPPIST-1 ephemeris with 
observations obtained at the location of the La Silla Observatory in Chile. 

The window functions within transit are sampled separately in order to control the number of 
observations used to measure the RM effect and constrain $\beta$. Due to the short transit widths 
of the two innermost planets ($\lesssim 42$ minutes) and the typical integration time of 
15 minutes plus overhead, in order 
to obtain well-sampled transit windows requires numerous transits in 
which at most two radial velocity 
measurements can be made. This can potentially introduce astrophysical noise 
in the RM timeseries as levels of stellar jitter may vary between transit events. 
The effect of the integration time on the amplitude of the RM effect is 
discussed in Sect.~\ref{sect:rmeffect}.

For each unique set of timeseries hyperparameters (i.e. $N_{\mathrm{obs}}$ both inside and outside 
of transit), we generate ten independent window functions 
to search for planets. In this way, we can 
marginalize over the hyperparameters when calculating the 
planet mass and $\beta$ measurement uncertainties. 

\subsection{Planetary Signals}
Radial velocity (RV) timeseries of the star with planetary companions will 
contain a number of components which 
may or may not have an astrophysical origin. The first 
component arises from the gravitational influence of the three TRAPPIST-1 planets reported  
in \citetalias{gillon16}. 
We model these contributions as the superposition of three keplerian orbits with the orbital 
periods shown in Table~\ref{trappist1table}. Although the orbital period of TRAPPIST-1d is not 
significantly well-constrained by the current data, we adopt the `most-likely' value from 
\citetalias{gillon16} 
but do not attempt to recover the mass or spin--orbit angle of this planet in 
our simulations. To compute planetary masses from 
their observed radii, we assume a bulk Earth-like density. This mass-radius relationship 
for planets with $r_p \gtrsim 1.6$ R$_{\oplus}$ is supported by the highest precision 
measurements 
of transiting planets \citep{dressing15b} and from theoretical planet formation studies of 
sub-Neptune sized objects \citep{lopez14}. Our simplified mass-radius relationship results 
in a conservative estimate of planetary masses compared to that which is obtained by 
other, empirically derived mass-radius parameterizations (\citeauthor{cloutier16} \emph{in prep}). 
The resulting planet masses are $m_p = 1.379, 1.154, 1.573$ M$_{\oplus}$ for TRAPPIST-1b, c, and d 
respectively. Following the analysis in \citetalias{gillon16}, 
each orbit is assumed circular resulting in Doppler semi-amplitudes of 
$K_{\mathrm{Dop}} = 4.14, 2.96, 2.06$ \ms{} for TRAPPIST-1b, c, and d respectively. 

One caveat of the keplerian model may arise from the compactness of the planetary 
system, particularly the two innermost planets, wherein mutual interactions between 
the planets may not be negligible despite the planets' small semimajor axes and 
therefore efficient tidal circularization. Any planet-planet interactions may be able to 
introduce small non-zero eccentricities. However any such eccentricities are expected to 
be small and we therefore ignore their effects in our analysis. Furthermore, 
we've declined the use of a fully dynamical model 
due to the planetary transit light curves not exhibiting any significant transit-timing 
or transit-duration variations \citepalias{gillon16} and 
thus justifying the adopted keplerian model as a realistic approximation.

Additional planets on wide orbits are posited to exist in this system 
which would certainly contribute to the observed radial velocity signal. 
However this prediction is not testable with the currently available data so we limit the 
number of dynamical components to the number of known planets in the TRAPPIST-1 system. 
Furthermore, we note that regarding Earth-like planets in other M-dwarfs systems,  
the average number of small planets per M-dwarf is $\sim 2.5$ \citep{dressing15a} so the 
inclusion of 3 planets in our synthetic timeseries are naturally representative of 
the dynamical contribution from small planets in typical M-dwarf systems. 

\subsection{Stellar Jitter} \label{sect:jitter}
Low mass stars undergoing rapid rotation, 
such as those required to exhibit a detectable RM effect from small planets, 
are known to often exhibit correspondingly large levels of stellar activity 
\citep{mohanty03, browning10, reiners12a, west15} whose signals will be manifested 
in the observed radial velocities. \citetalias{gillon16}  used the photometric light 
curve to measure the stellar rotation period of TRAPPIST-1 to be $\sim 1.4$ days. 
The quasi-periodic photometric variability is attributed to the rotation of 
active regions in the stellar photosphere such as spots and plages 
which contribute to the radial velocity signal in two distinct ways: 
i) as the active regions traverse the 
stellar disc they will disturb its axial symmetry, similar to the RM effect, thus causing 
a radial velocity variation that scales with the fractional surface coverage by 
active regions and its first derivative and ii) active regions 
will suppress the convective blueshift at the photospheric boundary 
\citep{aigrain12}. We find that the former is the dominant source of RV jitter 
from active regions 
for rapidly rotating stars such as TRAPPIST-1 and is referred to as the flux effect. 

An additional source of RV jitter arises from the Zeeman broadening of 
spectral features in stars with significant magnetic fields. \cite{reiners13} argued 
that the Zeeman RV jitter is proportional to the square of the magnetic field strength 
and increases towards longer wavelengths as the Zeeman displacement itself grows with 
wavelength. \cite{reiners10b} measured the magnetic field strength of TRAPPIST-1 to be 
600 Gauss. From this we compute the RV jitter from Zeeman broadening to be a strong 
contributor to the jitter but still with a smaller amplitude than the flux effect as a 
result of its relatively weak magnetic field strength.

Both the flux effect and Zeeman broadening dominate the RV jitter budget over the 
suppression of convective blueshift and each depend on the relative fraction of active 
region coverage over time. The fractional coverage by active regions is derived from 
the photometric light curve of TRAPPIST-1 \citepalias{gillon16}. 
Using the light curve to derive the total RV jitter, 
we find that TRAPPIST-1 exhibits a large jitter signal ($\sim 20$ m s$^{-1}$) 
compared to any gravitationally induced signals resulting from its rapid rotation. 
The RV jitter model is evaluated in the near-IR at J band ($\lambda \sim 1.2$ $\mu$m) 
throughout this work. The short rotation period of TRAPPIST-1 is typical among the subset of 
late M-dwarfs which are rapidly rotating. Hence, our fiducial case of TRAPPIST-1 is 
representative of the population of M-dwarf systems conducive to the detection of the RM 
effect from an Earth-like planet (see Sect.~\ref{sect:typical}).
 
The large level of RV jitter should be present for 
TRAPPIST-1 despite the small amplitude of the star's photometric variability and 
low activity levels \citep{schmidt07, lee10, reiners10b, gillon16}. Fortunately, 
due to the fact that the radial velocity 
jitter is modulated by active regions\footnote{We assume that the jitter induced by 
granulation (i.e. not associated with active regions) can be averaged down via long 
integration times \citep{lovis05}.} which evolve in a quasi-periodic manner, we can use the 
observed photometric variability to model the RV jitter and partially remove it from 
the observed radial velocities whilst preserving the signal induced by planetary companions 
which do not affect the star's light curve outside of their transit windows. 
We note an important assumption made in this critical step: in order to partially 
mitigate the 
stellar jitter we must assume that the star's activity properties (e.g. the 
fractional coverage, spatial distribution, and differential rotation of active regions) 
do not vary over many rotation cycles. Alternatively, one may obtain 
contemporaneous photometry and radial velocities to measure the fractional coverage 
by active regions in real 
time. Unfortunately this is often not feasible for these types of observations and 
observers are forced to use additional ancillary timeseries such as line indicators 
(e.g. $\log{R'_{\mathrm{HK}}}$) or the bi-sector inverse slope. 
The accuracy with which the stellar jitter can be removed using photometry 
is primarily determined 
by the photometric precision which we adopt from \citetalias{gillon16}. 
Using the formalism from \cite{aigrain12} to model the flux effect and the approximate 
scaling of the Zeeman broadening from \cite{reiners13}, we reduce the level of stellar 
jitter down to an rms 
of $\sim 4$ \ms{} from $\sim 20$ \ms{} which is comparable to the noise level of 
prescribed instrumental/systematic uncertainties (see Sect.~\ref{sect:nonastronoise}). 
The use of novel machine learning techniques such as Gaussian process regression 
to model the stellar jitter may be able improve the residual rms to even lower levels 
depending on the value of $\sigma_{\mathrm{RV}}$ 
(\citeauthor{cloutier16} \emph{in prep}). 

TRAPPIST-1, and other cool stars, can potentially exhibit stochastic 
flares in their light curves 
on a typical timescale of minutes to hours \citep{davenport14, gillon16}. 
\cite{reiners09} argued that these energetic flares are easily distinguishable in spectral 
orders containing prominent emission lines \citep[e.g. H$\alpha$, H$\delta$, with optical 
spectrographs, or with near-IR spectrographs, P$\gamma$, P$\delta$ in Y-band, 
and Br$\gamma$ in K-band;][]{schmidt12} which are excited 
beyond the stellar quiescent state during flaring events. 
Following \cite{reiners09} we assume that 
these events can be identified from the sudden presence of the aforementioned emission 
signatures and the corresponding radial velocity measurements can be excluded when 
modelling planets. 

\subsection{Rossiter-McLaughlin Effect} \label{sect:rmeffect}
The Rossiter-McLaughlin effect of the two inner-most planets is modelled using the 
equations from \cite{gimenez06}. The relevant stellar and planetary parameters from 
Table~\ref{trappist1table} are used to compute the models. Because transit widths 
for small planets around M-dwarfs are typically on the order of $\sim30-120$ minutes, 
the integration time required to measure the stellar radial velocity (typically 
15-20 minutes) occupies a significant fraction of the transit window. To account for 
this we smooth the RM model over bins with a width equal to the 15 minute integration 
time. This effect reduces the 
RM semi-amplitude to below its maximum theoretical value thus making the 
RM effect more difficult to detect.

Because the structure of the RM effect, and hence $K_{\mathrm{RM}}$, span a range 
of forms, the maximum value of $K_{\mathrm{RM}}$ is realized at $\beta=0^{\circ}$ and 
goes to zero as $|\beta| \to 90^{\circ}$. Therefore we separately consider 
four cases of the planetary spin--orbit angle $\beta=0^{\circ},30^{\circ},60^{\circ},90^{\circ}$ 
where the $\beta=0^{\circ}$ corresponds to a perfectly aligned orbit and 
$\beta=90^{\circ}$ corresponds to a planetary orbital plane parallel to the stellar 
spin axis wherein no RM effect is generated as the planet only occults the 
region of the stellar disc whose rotation velocity is purely tangential to the 
line-of-sight. We do not consider retrograde orbits by their waveform 
symmetry with the spin--orbit angles sampled. 
Because the empirical distribution of mutual inclinations for small 
planets around M-dwarfs shows that these systems favour co-planarity 
\citep{figueira12, fabrycky14}, we assign 
a common value of $\beta$ to each planet in each simulation. 

\subsection{Non-Astrophysical Noise} \label{sect:nonastronoise}
Finally we add an additional noise model which is not astrophysical in nature as it is 
determined by the experiment's instrumentation. Explicitly we add a conservative white noise 
contribution at the level of 6 \ms{} akin to the level of RV precision demonstrated by 
near-IR velocimeters so far. To address cases of the next generation of near-IR 
velocimeters 
(e.g. \emph{SPIRou}; \citealt{thibault12, artigau14}, \emph{CARMENES}; 
\citealt{quirrenbach14}, \emph{HPF} \citealt{mahadevan12}, and \emph{IRD} spectrograph 
\citealt{tamura12}) we also consider an improved level of RV precision of 2 \ms{.} 
In addition we add a systematic noise term 
to account for any non-Gaussian noise and allows us to vary the global noise properties of 
the radial velocity timeseries in a simple parametric way. 

\begin{figure*}
\centering
\includegraphics[scale=0.4]{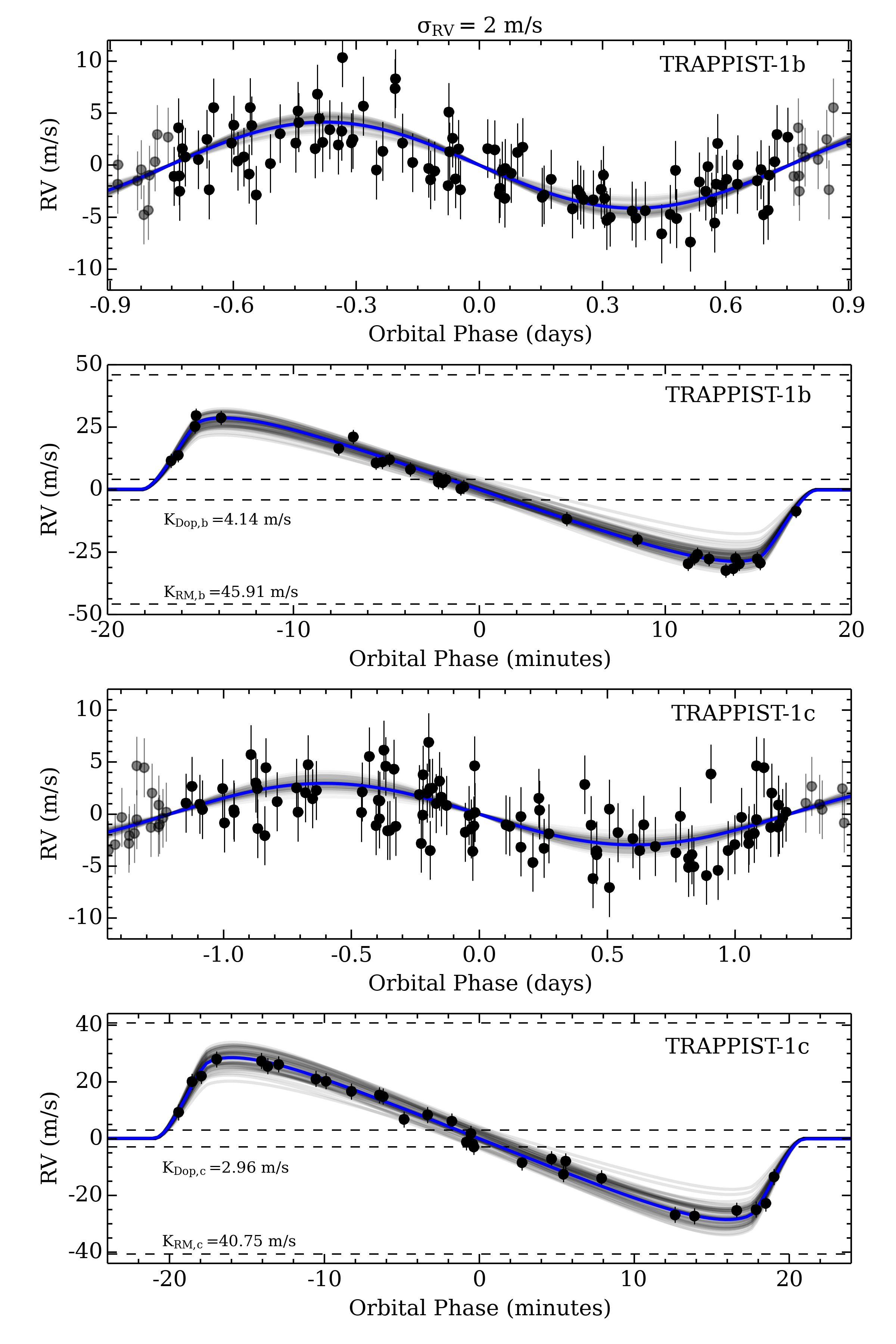}
\includegraphics[scale=0.4]{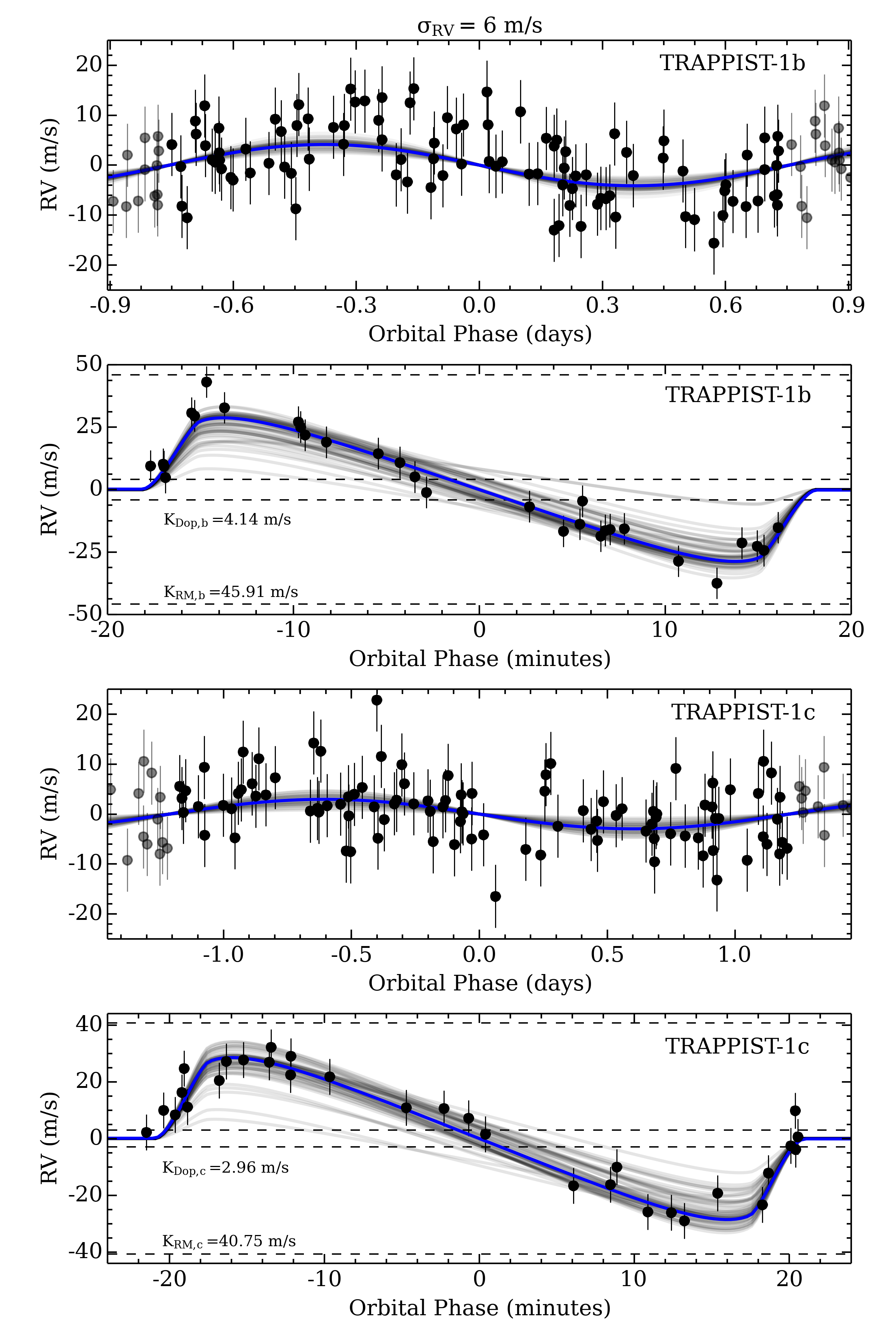}
\caption{Sample phase-folded, simulated radial velocity 
timeseries for TRAPPIST-1b and c with 
$\sigma_{\mathrm{RV}}=2$ \ms{} (\emph{left panels}) and 
$\sigma_{\mathrm{RV}}=6$ \ms{} (\emph{right panels}). 
Directly beneath each full RV timeseries is a close-up view of the transit window 
revealing the Rossiter-McLaughlin effect. 
In all panels the \emph{true} model from which the data are derived is shown 
in \emph{blue}. Models computed from random draws from the model parameter 
posterior distributions are shown in \emph{translucent grey}. The \emph{horizontal 
dashed lines} in the RM panels illustrate the Doppler semi-amplitude of the 
corresponding planet and the semi-amplitude of the Rossiter-McLaughlin effect in 
the absence of smoothing the observations over the integration time. 
For clarity, the full RV panels do not contain any measurements 
made in-transit while the RM panels do not contain measurements made outside of 
transit. \label{fig:timeseries}}
\end{figure*}

For reference, examples of our timeseries for TRAPPIST-1b and c 
are shown in Fig.~\ref{fig:timeseries} for the two values of 
RV precision considered. Each timeseries is phase-folded to the planet's orbital period 
and has the keplerian models from the other two planets removed. For clarity, 
measurements made in-transit are excluded from plots of the full RV timeseries because 
their values can be an order of magnitude greater than the planet's Doppler 
semi-amplitude. The large amplitude RM waveforms are then shown below each full RV 
timeseries for the $\beta=0^{\circ}$ case. In these RM panels, measurements made outside 
of the transit window are excluded. 

\section{Planet Model Fitting} \label{model}
As we are interested in the recovery of the RM effect due to the two innermost planets, 
we must first obtain the best-fit keplerian model to remove from the dataset and search 
for the RM signature in the residuals. We use the \texttt{emcee} \citep{foremanmackey13} 
Markov chain Monte-Carlo ensemble sampler to recover the keplerian model parameter posterior 
probability density functions (PDF). Because the planetary orbital periods and times of 
mid-transit are so well constrained by their transit light curves, we take their values to 
be absolute. 
Hence we are left with just three free parameters, namely, the RV semi-amplitudes of each of 
the three planets. This limits the extent of the model parameter posterior probability space 
and decreases the MCMC's wall time. 

For each synthetic RV timeseries (see Sect.~\ref{methods}) we allocate 100 walkers to 
explore the model parameter space. Each walker is initialized in a Gaussian ball 
around the approximate `true' model parameter values. The walkers act as correlated Markov 
chains, each with a burn-in length defined such that the chains run for $\sim 9-12$ 
autocorrelation times to ensure that we obtain the equivalent number of independent samples 
of the target 
parameter posterior PDFs. Following the burn-in phase, the walkers are extended to another  
9-12 autocorrelation times in order to compute the model posterior PDFs and constrain the 
measurement uncertainty on each parameter. Furthermore, the mean acceptance fraction among 
the walkers is monitored throughout the MCMC simulations and modifications to the walker 
initialization, or seldomly to the acceptance fraction scale parameter\footnote{The scale 
parameter can be used to tune the probability that a proposed step in each walker's chain 
is accepted or rejected. See Sect. 4 of  \cite{foremanmackey13}}, 
are used to constrain the mean acceptance fraction to an ideal range of $\sim 20-50$\%. 

Given the estimates of TRAPPIST-1 planet masses and their measured orbital periods, we are 
confident that their RV semi-amplitudes are $< 10$ \ms{}. From this we adopt a common, 
uniform prior on each planet's 
semi-amplitude; $K \in \mathcal{U}(0,10)$ \ms{}. Furthermore we adopt the 
Gaussian likelihood function proposed by \cite{gregory10} in our MCMC. 

Following the recovery of the maximum \emph{a posteriori} (MAP) RV semi-amplitudes, we couple 
those parameter values to the planetary orbital parameters and compute the MAP keplerian 
models. These models are removed from the timeseries and we proceed to search for the RM 
waveform in the residuals. Specifically we focus on the time interval spanning one transit 
duration prior to transit ingress and following transit egress ($\approx 100$ minutes and 
$\approx 125$ minutes for TRAPPIST-1b and c respectively). 

Fixing most planetary and stellar model parameters in the \cite{gimenez06} models of the 
RM effect, we set the free parameters in the MCMC to \sqvsini{$\cos{\beta}$} and 
\sqvsini{$\sin{\beta}$}. This prescription follows from \cite{triaud11} for the purpose of 
reducing correlations among the free model parameters. The MCMC is then run identically as 
above to obtain independent measurements of \vsini{} and the spin--orbit angle 
$\beta$. The limits on our adopted uniform 
prior on \vsini{} are set by the fact that \vsini{} must be $> 0$ km s$^{-1}$ 
for non-zero inclination\footnote{It is expected that $i_s \neq 0^{\circ}$ 
given that periodic modulations are observed 
in the TRAPPIST-1 light curve \citepalias{gillon16}.} and the $3\sigma$  
uncertainty on \vsini{} from spectroscopic measurements. 
Hence, \vsini{} $\in \mathcal{U}(0,12)$ km s$^{-1}$. As we have 
effectively no \emph{a priori} knowledge of the planetary spin-angles in the 
system\footnote{Aside from the empirical evidence that small planets 
rarely undergo high eccentricity migration \citep[e.g.][]{figueira12, fabrycky14} 
therefore favouring well-aligned orbits; 
$\beta \sim 0^{\circ}$.}, we must consider the full range of possible spin--orbit angles: 
$\beta \in \mathcal{U}(-\pi, \pi)$. 

For some test cases with various input spin--orbit angles, we compared the results 
of our MCMC analysis of the RM effect when using the aforementioned uniform prior  
to the use of the Jeffreys non-informative prior. We found that the resulting MAP \vsini{} 
and $\beta$ values and their uncertainties were effectively unchanged by our choice of 
prior.

\section{Results} \label{results}
\subsection{Recovery of Planet Masses} \label{sect:massdet}
Before we can investigate the measurement precision of the spin--orbit angle $\beta$, we 
require an orbital solution for the system. The best-fit orbital solution from MCMC will be 
removed from the radial velocity timeseries which allows one to search 
the residuals for the anomalous RM effect within each of the two inner planets' known transit 
windows. Any discrepancy between the best-fit orbital solution and the `true' planetary 
orbits will introduce an additional source of error into the residual timeseries and affect 
the accuracy with which the spin--orbit angle can be measured.

\begin{figure}
\centering
\includegraphics[scale=0.5]{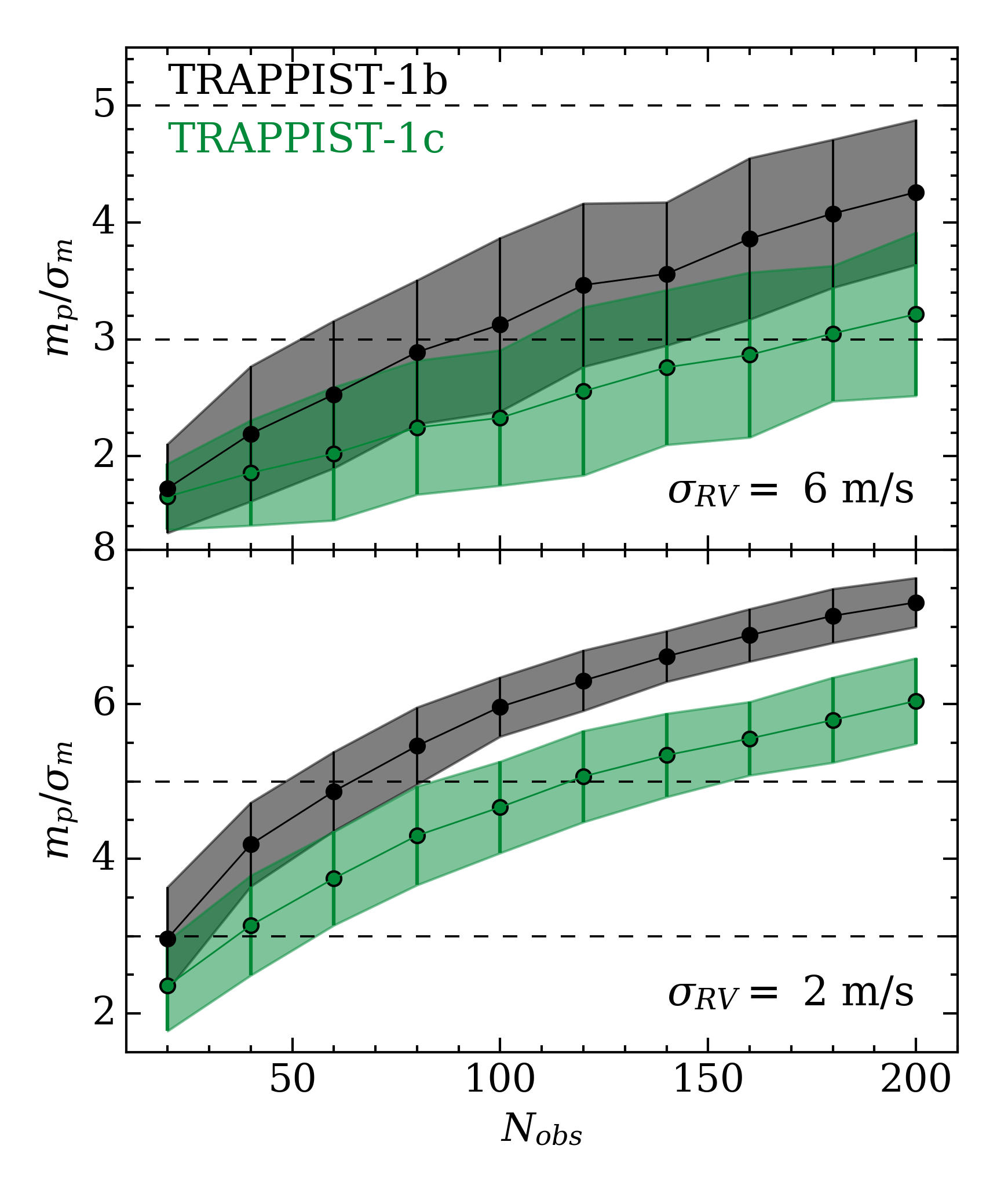}
\caption{The mass detection significance, $m_p / \sigma_{m_p}$, as a function of the 
number of the radial velocity measurements used to fit the Doppler semi-amplitudes, 
for two values of the fixed radial velocity uncertainty; $\sigma_{\mathrm{RV}}=2$ m s$^{-1}$ 
(\emph{top}) and $\sigma_{\mathrm{RV}}=6$ m s$^{-1}$ (\emph{bottom}). The shaded regions 
highlight the approximate $1\sigma$ confidence intervals for each planet. \emph{Horizontal 
dashed lines} are used to indicate a 3 and $5\sigma$ mass detection in each panel. 
\label{fig:massdet}}
\end{figure}

A product of the first set of MCMC simulations from Sect.~\ref{model} is the posterior PDFs 
of the planet keplerian semi-amplitude solutions $K_i$ for $i=$ b,c,d. 
We take the parameter uncertainties to be 
16$^{\mathrm{th}}$ and 84$^{\mathrm{th}}$ percentiles ($1\sigma$ if the PDF is Gaussian) of the 
posterior PDF and the best-fit value to be the MAP value. Knowledge of the $K_i$ 
uncertainties are crucial as they will ultimately affect the precision of our best-fit 
orbital model and therefore our ability to accurately measure $\beta$. 
Using the MCMC recovered values of $K_i$, the orbital period, orbital 
inclination, and stellar mass from Table~\ref{trappist1table}, 
we use the standard method of error propagation to compute the MAP planet mass $m_p$ and 
its associated uncertainty $\sigma_{m_p}$ from each synthetic RV timeseries. We then define 
the mass detection significance as $m_p / \sigma_{m_p}$. Fig.~\ref{fig:massdet} shows 
the median mass detection significance for each of the two inner planets as a function of 
the number of RV measurements used to fit the model. The uncertainty in each \nobs{} bin 
is calculated from the median absolute deviation of the mass detection significances within 
that bin.

As expected, the mass detection significance increases monotonically with \nobs{} and 
appears to asymptotically approach a maximum mass detection significance which  
requires additional simulations with \nobs{} $> 200$ to measure. Furthermore, the mass 
detection significance of TRAPPIST-1b is always greater than for TRAPPIST-1c due to 
its larger Doppler semi-amplitude. 

Considering the $\sigma_{\mathrm{RV}}$ cases separately, unsurprisingly we see that it has a 
significant effect on the mass detection significance for a given number of RV measurements. 
In particular, for $\sigma_{\mathrm{RV}}=2$ \ms{} a $3\sigma$ mass detection for  
TRAPPIST-1b or c is achieved with $\sim 20$ and 40 measurements respectively and 
for TRAPPIST-1b increases to a 
$\geq 5\sigma$ detection with \nobs{} $\gtrsim 60$. TRAPPIST-1c requires 
\nobs{} $\sim 120$ in order to achieve a $5\sigma$. 
These results for $\sigma_{\mathrm{RV}}=2$ \ms{} 
are broadly consistent with the mass detection significance of GJ 1132b 
\citep{berta15}, a rocky planet transiting a nearby mid-M-dwarf (\citeauthor{cloutier16}  
\emph{in prep}). If the RV measurement uncertainty is degraded to 
$\sigma_{\mathrm{RV}}=6$ m s$^{-1}$, the mass detection significance is reduced. Namely, 
a $3\sigma$ mass detection of TRAPPIST-1b (c) requires \nobs{} $\sim 90$ ($\sim 180$). 
Furthermore, even 
with 200 measurements, the mass detection significance is always $< 5\sigma$ for either 
planet. Clearly, a RV uncertainty of 6 \ms{} compared to 2 \ms{} significantly 
deters one's ability to detect small transiting planet masses in RV with a modest number 
of RV measurements (\nobs{} $\lesssim 100$).

\subsection{Detecting the RM effect and Spin--Orbit Angle Measurements} \label{sect:beta}
\begin{figure*}
\centering
\includegraphics[scale=0.6]{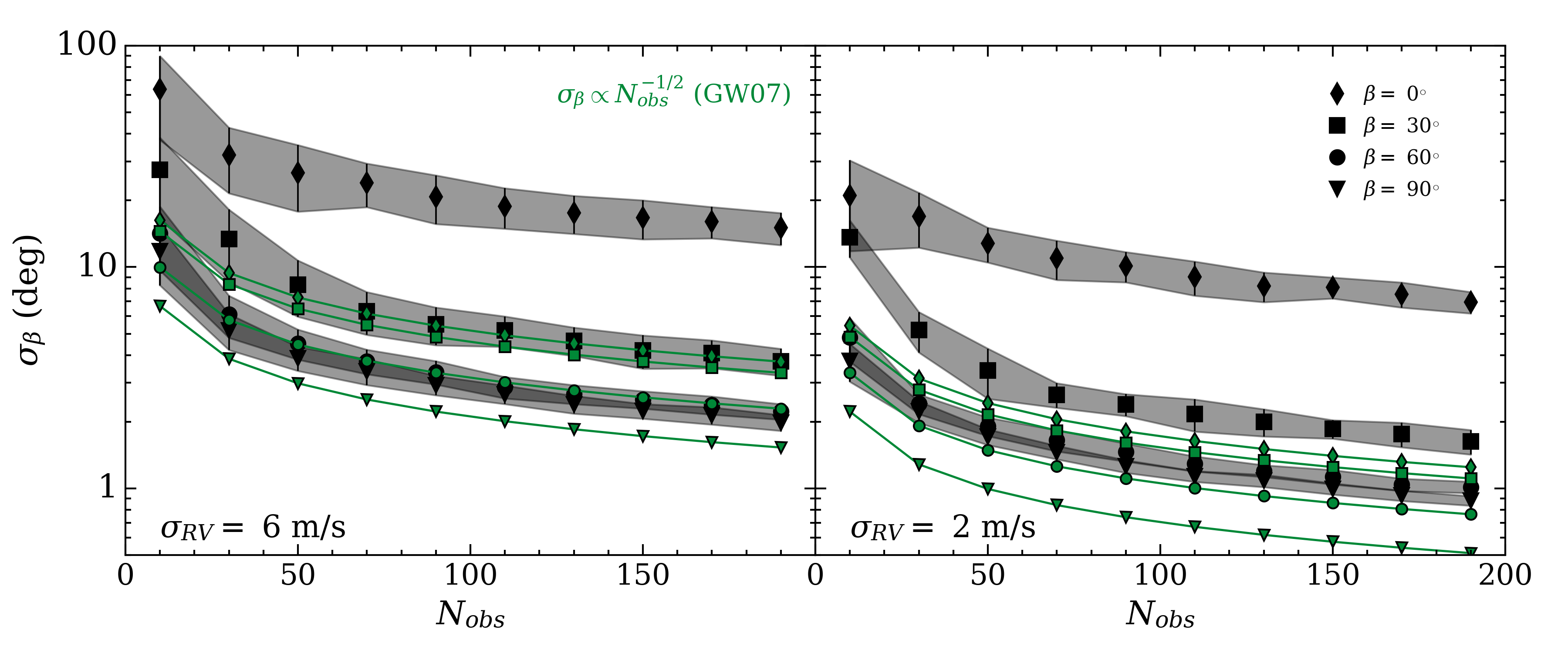}
\caption{The average measurement uncertainty of the spin--orbit angle for TRAPPIST-1b and 
c from the measurement of the Rossiter-McLaughlin waveform as a function of the number of 
radial velocity measurements made in-transit. Values of 
$\beta=0^{\circ},30^{\circ},60^{\circ},90^{\circ}$ are considered for both   
cases of the radial velocity measurement uncertainties ($\sigma_{\mathrm{RV}}=6$ 
m s$^{-1}$; \emph{left} and $\sigma_{\mathrm{RV}}=2$ m s$^{-1}$; \emph{right}). The shaded regions 
approximately depict the $1 \sigma$ confidence intervals from the dispersion in $\sigma_{\beta}$ after 
multiple Monte-Carlo realizations. The \emph{green curves} represent 
the predicted measurement uncertainty from \citetalias{gaudi07}.} 
\label{fig:betaerr}
\end{figure*}

Using the keplerian model solutions described in Sect.~\ref{sect:massdet}, we are 
in a position to model the RM effect in the RV residuals. Given the unique opportunity 
to detect the RM effect of Earth-like planets around small stars, we are interested 
in quantifying the spin--orbit angle measurement precision achievable. 
Fig.~\ref{fig:betaerr} shows the average measurement uncertainty of the spin--orbit 
angle $\sigma_{\mathrm{\beta}}$ 
between the two innermost planets for $\sigma_{\mathrm{RV}}=2$ and 6 \ms{.} 
Only the RV measurements made within the planet's transit window are used to 
fit the RM waveform and measure $\beta$. 

As was seen with the planet mass measurements 
in Fig.~\ref{fig:massdet}, the spin--orbit angle measurement uncertainty in 
Fig.~\ref{fig:betaerr} decreases monotonically with \nobs{.} At a given value 
of \nobs{,} the absolute value of $\sigma_{\beta}$ decreases for 
increasing $\beta$ from $0^{\circ}$ to $90^{\circ}$. This is 
due to the general trend of decreasing $K_{\mathrm{RM}}$ as $|\beta| \to 90^{\circ}$ 
for the impact parameters of the TRAPPIST-1 planets 
since the dispersion of the RM waveform about the radial velocity 
zeropoint is minimized for $\beta=90^{\circ}$ 
implying that as the structure in the waveform 
increases, more values of $\beta$ become consistent with the data at a given 
$\sigma_{\mathrm{RV}}$ causing the measurement uncertainty to grow.

Although we do not yet have a statistically significant sample of spin--orbit angles of 
Earth-like planets, their observed low mutual inclinations in multi-planet systems 
suggests that their orbital alignments will be consistent with $\beta = 0^{\circ}$ as a 
result of their cold dynamical histories \citep{figueira12, fabrycky14}. 
Focusing on this well-aligned 
case, it is apparent that $\sigma_{\beta}$ will be large ($\sigma_{\beta} > 10^{\circ}$) 
even up to \nobs{} $=190$ if $\sigma_{\mathrm{RV}}$ is large (6 \ms{)}.\footnote{We 
note that obtaining such a large number of in-transit 
measurements, with at most two measurements per observed transit, is difficult 
in practise as the timescale for doing so likely exceeds that of a 
single observing season.}
In order for $\sigma_{\beta}$ to be $<20^{\circ}$, \nobs{} must be $\gtrsim 110$. 
This improves dramatically 
if $\sigma_{\mathrm{RV}}$ is reduced to 2 \ms{} with $\sigma_{\beta}<20^{\circ}$ when 
\nobs{} $\gtrsim 10$ and $\sigma_{\beta} \sim 7^{\circ}$ at \nobs{} $=190$. 
Furthermore, each of the $\beta = 0^{\circ}$ curves shown in Fig.~\ref{fig:betaerr} 
appear to approach a minimum value of $\sigma_{\beta}$ close to the aforementioned 
values implying that additional RV measurements do not drastically improve your 
$\beta$ measurement precision. In practice, $\sigma_{\beta} \sim 7^{\circ}$ 
($\sigma_{\beta} \sim 16^{\circ}$) for $\sigma_{\mathrm{RV}} = 2$ \ms{} 
($\sigma_{\mathrm{RV}} = 6$ \ms{)} provides 
the most stringent precision with which we can measure $\beta$ unless many hundreds 
more observations can be obtained or the RV rms can be decreased either by more accurate 
modelling of the stellar jitter and/or by improving the instrument's RV stability 
(\citeauthor{cloutier16} \emph{in prep}). 

Complimentary to our empirical estimate of the spin--orbit angle measurement 
uncertainty, \citep[][hereafter \citetalias{gaudi07}]{gaudi07} 
presented an analytical calculation of $\sigma_{\beta}$ which 
differs from our results but are fairly consistent with certain observational results 
of hot Jupiters \citep[e.g.][]{triaud10}. We  
attribute the discrepancies between the two studies to differences in the noise properties 
of the observed timeseries. Unlike in \citetalias{gaudi07}, our RV timeseries include 
significant non-white noise contributions 
from stellar jitter (see Sect.~\ref{sect:jitter}) that are partially removed, 
leaving behind RV residuals that differ substantially from white. This is not the 
case in the test cases presented in \citetalias{gaudi07} or in the observational studies which 
roughly agree with the \citetalias{gaudi07} estimation of $\sigma_{\beta}(N_{\mathrm{obs}})$. This is because those 
systems containing hot Jupiters exhibit planetary 
signals that are large compared to the stellar jitter unlike the case of Earth-like 
planets around rapidly rotating, small stars.

The analytic expression of \citetalias{gaudi07} 
($\sigma_{\beta} \propto N_{\mathrm{obs}}^{-1/2}$; see their Eq. 16) 
is over-plotted in Fig.~\ref{fig:betaerr}. This expression underestimates the spin--orbit 
angle measurement uncertainty for each value of $\beta$ as a result of the dominant noise properties 
being non-white. In addition to differences in the noise properties, 
discrepancies are derived from the differences in their 
respective methodologies. The 
expression in \citetalias{gaudi07} is derived from the Fisher information 
in those systems and an analytical prescription of the RM waveform under the assumption of 
large \nobs{} and measurements which are uniformly sampled in time. 
Conversely, our method of deriving 
$\sigma_{\beta}$ is intended to mimic that of a realistic observing procedure 
wherein a sampling method (e.g. MCMC) is used to compute the RM model parameter posterior 
probability distributions from which measurement uncertainties are derived. In this 
way our methodology is independent of the approximations made when deriving an analytical 
approximation and is representative of how model parameter uncertainties are derived in 
practise. Indeed this is the procedure used in a number of observational studies of the RM 
effect \citep[e.g.][]{queloz10, triaud10}.

\section{Discussion} \label{discussion}
\subsection{Constraints on Planetary System Alignment}
If the RM effect is detected for each of TRAPPIST-1b and c, then we can compute the 
spin--orbit angle difference $\Delta \beta = \beta_b - \beta_c$ and its 
corresponding precision. This quantity is useful for determining whether or not the 
planetary system is mutually aligned. For instance, a value of $\Delta \beta \neq 0$ 
provides evidence for independent dynamical histories among the two planets. 
Recall that during the construction of the RM timeseries, 
$\Delta \beta = 0^{\circ}$ was imposed in all cases. 
From measurements of $\beta_b$ and $\beta_c$ and their uncertainties (see 
Fig.~\ref{fig:betaerr}) the uncertainty on $\Delta \beta$ can easily be computed via 
the product of $\sigma_{\beta}$ with $\sqrt{2}$ from the propagation of errors 
as the uncertainty in $\beta_b$ and $\beta_c$ should be equal when 
$\Delta \beta= 0^{\circ}$. 

To summarize the precision with which we can measure $\Delta \beta$, we focus on the 
case with $\beta=0^{\circ}$, the expected mean result for small planets around ultracool 
dwarfs. When $\sigma_{\mathrm{RV}} =6$ \ms{,} $\sigma_{\Delta \beta}$ decreases from 
$\sim 90^{\circ}$ for \nobs{} $=10$ down to $\sim 21^{\circ}$ for \nobs{} $=190$. 
Comparatively, when $\sigma_{\mathrm{RV}} =2$ \ms{,} $\sigma_{\Delta \beta}$ decreases 
from $\sim 30^{\circ}$ for \nobs{} $=10$ down to $\sim 10^{\circ}$ for \nobs{} $=190$. 
Of course these values will be smaller for larger values of $\beta$ but we leave 
interested readers to estimate $\sigma_{\Delta \beta}$ 
for themselves using the $\sigma_{\beta}$ curves shown in Fig.~\ref{fig:betaerr}.
For example, for a `reasonable' value of \nobs{} $=50$ and 
$\sigma_{\mathrm{RV}}=2$ \ms{,} we can measure the misalignment of the planetary orbital 
planes to $\pm 17^{\circ}$. This becomes $\pm 40^{\circ}$ if $\sigma_{\mathrm{RV}}=6$ \ms{.}

\subsection{Atmospheric Studies}
The presence of various absorbing species within the atmospheres of exoplanets 
can be measured using transmission spectroscopy as a variation in the 
apparent size of the planet with wavelength \citep{seager00}. 
\cite{snellen04} and \cite{digloria15} managed 
to detect the chromatic RM effect in the atmospheres of the hot Jupiters 
HD 209458b and HD 189733b respectively. 
Although \cite{dewit16} recently demonstrated that neither 
TRAPPIST-1b nor c possess a hydrogen/helium dominated atmosphere, recent 
theoretical studies \citep[e.g.][]{owen16} have shown that such an atmospheric 
composition still remains a possibility for planets similar to those 
orbiting TRAPPIST-1 and are expected to be detected with upcoming instrumentation 
\citep[e.g. \emph{SPECULOOS};][]{gillon13}.

For the sake of argument if we adopt TRAPPIST-1b as a fiducial case but include a 
primarily hydrogen/helium atmospheric composition ($\mu=2.5$) then one would expect 
a large pressure scale-height of $\sim 110$ km with an assumed Bond albedo of 0.3 
($T_{\mathrm{eff}} = 365$ K). In the absence of clouds, spectral features from 
near-IR absorbing species (e.g. H$_2$O and CH$_4$) mixed into the primarily H/He atmosphere 
can reach 6-10 scale heights \citep{millerricci09, dewit13}. Taking this lower factor 
applicable to the Earth \citep{kaltenegger09}, the expected variations in the 
apparent size of the planet will be of order 1500 ppm in spectral orders containing  
features from the absorbing species. This increase in opacity 
translates into a variation 
in the RM effect semi-amplitude of $\Delta K_{\mathrm{RM}} \sim 9$ m s$^{-1}$. 
Here \vsini{} amplifies the atmospheric signal.

For a `white-light' RM semi-amplitude 
of 45.9 m s$^{-1}$, an increase of 9 m s$^{-1}$ is a significant increase. 
From our simulations we found that so long as the RM waveform is well-sampled 
around its peak velocity, the uncertainty in the projected stellar rotation velocity 
is always $\lesssim 5$ km s$^{-1}$. 
This implies that the planetary atmosphere may be measurable with the RM effect at 
multiple wavelengths and is certainly detectable if indeed the planet's atmospheric 
mean molecular weight is dominated by hydrogen and helium and the signal-to-noise 
within the reduced wavelength bins can remain sufficiently high. 

\subsection{The Rossiter-McLaughlin Effect Around Late-type Stars}
\begin{figure}
\centering
\includegraphics[scale=0.5]{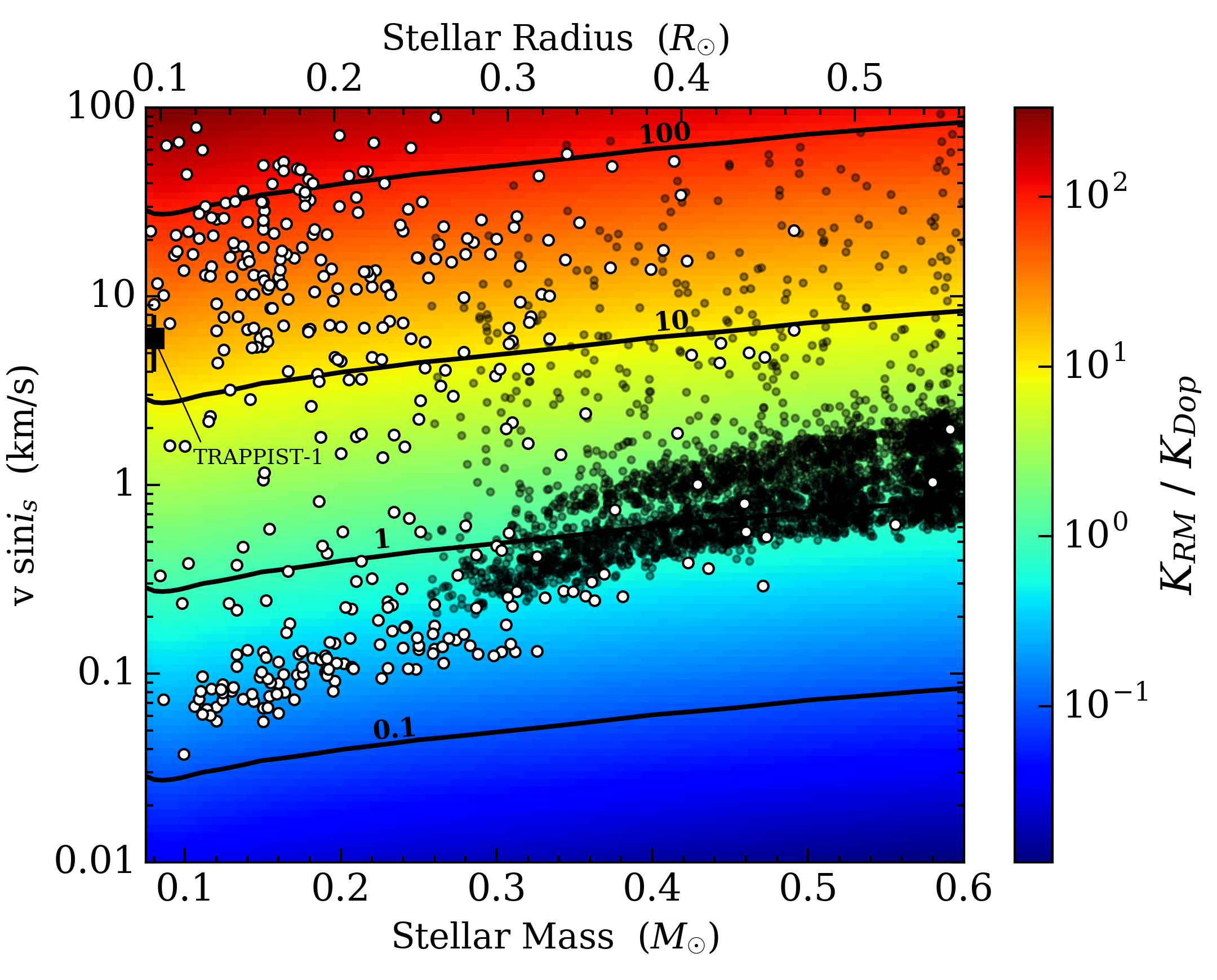}
\caption{The ratio of the Rossiter-McLaughlin semi-amplitude to the Doppler 
semi-amplitude for an Earth-sized planet at the 
inner edge of the habitable zone. \emph{White} and \emph{black} 
symbols depict stellar rotation periods measured with MEarth 
\citep{irwin11, newton16a} and Kepler \citep{mcquillan14} respectively. 
TRAPPIST-1 is depicted as the \emph{black square} according to its mass. 
Over the range of 
observed M-dwarf rotation velocities, the semi-amplitude of the 
RM effect can be over two orders-of-magnitude less than or greater than 
the Doppler semi-amplitude of a typical rocky planet in the habitable zone. 
\label{fig:Kratio}}
\end{figure}

The RM effect and hence spin--orbit angles of Earth-like planets has not been well-studied 
compared to the giant planets. This comes as a result of their often small transit depths and 
low RM semi-amplitudes. But given the large frequency of small planets around late-type stars 
and the large subset of stars which are rapidly rotating, there exists many targets for 
which the RM semi-amplitude of an Earth-like planet may be large compared to the planet's 
Doppler semi-amplitude and potentially two orders of magnitude greater. This is evidenced 
in Fig.~\ref{fig:Kratio} which depicts the $K_{\mathrm{RM}} / K_{\mathrm{Dop}}$ ratio as a 
function of stellar mass and projected stellar 
rotation velocity for an Earth-sized planet at the 
inner edge of the habitable zone; one of the most common types of planet around M-dwarfs 
\citep{dressing15a}. The inner edge of the habitable zone is computed at each grid 
point from the \cite{kopparapu13} prescription and the stellar effective temperature 
derived from the stellar mass at 5 Gyrs assuming solar metallicity \citep{baraffe98}, 
applicable down to 0.075 M$_{\odot}$. 
The M-dwarf rotation period distribution, converted to \vsini{} assuming $i_s=90^{\circ}$ 
and deriving the stellar radius from the stellar mass, 
is over-plotted and shows that a large population of M-dwarfs exists for which 
$K_{\mathrm{RM}} / K_{\mathrm{Dop}} \gg 1$ and are therefore prime targets for the RM 
characterization of an Earth-like planet. 

\subsection{How typical is the TRAPPIST-1 system?} \label{sect:typical}
It attempting to characterize the effort required to recover the RM effect due to 
Earth-like planets around the coolest stars, we have focused our attention on a particular 
system, namely TRAPPIST-1. Justification of this approximation should be emphasized. 
Firstly, the consideration of a single system greatly simplifies the computational 
cost of our study because the parameter space is substantially reduced. 
Furthermore, we argue that TRAPPIST-1 is a star that is well representative of 
the ultracool stellar population of interest for the detection of the RM effect 
from Earth-like planets. Notably, TRAPPIST-1 exhibits a rotation period of 
1.4 days from its photometric variability. \cite{irwin11} and 
\cite{newton16a} showed empirically that this is a typical value among 
late M-dwarfs over a wide range of ages as these stars 
can maintain short rotation periods well into their lifetimes \citep{west15}. 

In the limit of small transit 
depths, such as those typical of small planets around M-dwarfs, $K_{\mathrm{RM}}$ 
approximately scales linearly with transit depth \citep{gaudi07} such that 
$K_{\mathrm{RM}} \propto R_s^{-2}$. Because of this, detection of the RM effect 
favours small stars such as TRAPPIST-1 with $R_s=0.117$ R$_{\odot}$. However 
we argue below that the size of TRAPPIST-1 is representative of the population 
of stars close to the stellar/substellar boundary 
($0.075 \lesssim M_s \lesssim 0.1$). 

Because the MS lifetime of stars close to the stellar/substellar boundary 
exceeds the Hubble time, the typical radius of such stars can be 
estimated from the stellar IMF in that mass regime \citep{padoan02} 
coupled with stellar evolutionary tracks applicable to low-mass stars to 
compute the corresponding stellar radii \citep{baraffe98}. The analytical 
\cite{padoan02} IMF increases over this range of stellar masses before peaking 
around 0.2 M$_{\odot}$. Therefore the most common ultracool dwarf is approximately 
0.1 M$_{\odot}$ which has a model radius of $\sim 0.122 \pm 0.01$ R$_{\odot}$ 
by considering the effect of changing metallicity and relative He abundances. 
This value is consistent with the reported radius of TRAPPIST-1 (0.117 R$_{\odot}$). 
Therefore, the radius of TRAPPIST-1 is characteristic of the joint population 
of very low mass stars and brown dwarfs.

The planetary parameters of the TRAPPIST-1 planets 
are also typical of the population of small planets 
around late K to early M-dwarfs which we assume extrapolates down towards ultracool 
dwarfs given the lack observational constraints on the planet occurrence rate around 
ultracool dwarfs; a population which will be probed by the next generation of 
high precision near-IR spectrographs \citep[i.e. \emph{SPIRou};][]{thibault12, artigau14}. 
This extrapolation may be reasonable given the high number of small planets 
predicted to form around these cool stars via core-accretion \citep{payne07}. 

\subsection{Application to Other Systems}
The detection of the RM effect for Earth-like planets will be facilitated by the detection 
of such planets around the smallest stars including late M and brown dwarfs. The detection 
of the TRAPPIST-1 multi-planetary system in the TRAPPIST telescope science verification 
runs with just 50 targets, suggests that small planets are common around the smallest 
stars. This region of the stellar parameter space has been largely unexplored at present 
but will be investigated in the near future with dedicated transit 
\citep[e.g. \emph{SPECULOOS};][]{gillon13} and radial velocity surveys in the near-IR 
(e.g. \emph{SPIRou}; \citealt{thibault12, artigau14}, \emph{CARMENES}; 
\citealt{quirrenbach14}, \emph{HPF} \citealt{mahadevan12}, and \emph{IRD} spectrograph 
\citealt{tamura12}). The detection significance curves 
presented in Figs.~\ref{fig:massdet} and \ref{fig:betaerr} will help to inform the 
search for the RM effect from Earth-like planets around rapidly rotating small stars 
which will ultimately result in the direct measurement of 
the distribution of small planet spin--orbit angles; a quantity which 
is useful for probing the formation pathways of Earth-like planets around small stars. 

\section{Summary \& Conclusion} \label{conclusion}
In this study we have shown that the Rossiter-McLaughlin effect due to Earth-like planets 
is measurable around the class of ultracool dwarfs which includes late M to brown dwarfs. 
Such measurements are facilitated by the small stellar radii and high fraction of stars 
undergoing rapid rotation. 
Taking the TRAPPIST-1 multi-planetary system as a fiducial test case, we compute the 
detection significance of planetary masses and projected 
spin--orbit angles, via the Rossiter-McLaughlin 
effect, by constructing synthetic radial velocity timeseries and using Monte-Carlo 
simulations to measure the planet parameters of interest and their uncertainties. 
The main results are summarized as follows:

\begin{itemize}
\item Adopting a radial velocity stability of $\sigma_{\mathrm{RV}}=6$ \ms{}, a value 
representative of currently operating near-IR velocimeters, the planet masses of 
TRAPPIST-1b and c can be detected at the $3\sigma$ level with $\sim 90$ and 180 
measurements respectively. 
\item Decreasing the radial velocity stability to a more sought after level of 
$\sigma_{\mathrm{RV}}=2$ \ms{,} $3\sigma$ detections of the planet masses can be 
obtained with just $\sim 20$ and 40 measurements respectively.
\item Considering the probable cold dynamical histories of small planets around 
small stars, a projected 
spin--orbit angle measurement precision of $\sigma_{\beta} < 20^{\circ}$ 
requires \nobs{} $\gtrsim 110$ if $\sigma_{\mathrm{RV}}=6$ \ms{.}
\item If $\sigma_{\mathrm{RV}}=2$ \ms{,} this is greatly reduced to just \nobs{} 
$\sim 10$ and begins to asymptotically approach a minimum value of 
$\sigma_{\beta} \sim 7^{\circ}$. 
\item The significant non-white noise in radial velocity timeseries containing small 
planets around small, rapidly rotating stars (i.e. significant stellar jitter) 
causes $\sigma_{\beta}$ to be underestimated 
by analytical approximations derived in the limit of white noise domination \citep[e.g.][]{gaudi07}.
\item The resulting uncertainties imply that we cannot measure the difference in 
projected planetary spin--orbit angles between planets to better than $\sim 10^{\circ}$ 
when $\beta=0^{\circ}$. 
\item The sub-population of rapidly rotating M-dwarfs provides a set of targets 
for which $K_{\mathrm{RM}} \gg K_{\mathrm{Dop}}$ and represents the optimal targets for 
the direct measurement of the projected 
spin--orbit angle distribution for Earth-like planets. 
\item Variations in the planetary radius due to the presence of various volatiles 
is shown to result in a change in the TRAPPIST-1b transit depth of $\sim 1500$ ppm for 
a H/He dominated atmosphere. This corresponds to a measurable change in the 
Rossiter-McLaughlin semi-amplitude of $\Delta K_{\mathrm{RM}} \sim 9$ \ms{.}
\end{itemize}

Ultimately we are interested in the effort required to measure the projected spin--orbit 
angle of Earth-like planets via the detection of their Rossiter-McLaughlin effects. 
The TRAPPIST-1 planetary system represents a superlative target for such observations 
given the small size of the star and its rapid rotation velocity. Up-coming high 
precision, near-IR velocimeters will provide the best possible tools for making 
such measurements which will benefit greatly from achieving the smallest possible 
radial velocity stability. In the coming years, uncovering the planet population 
around the coolest stars and detecting their projected spin--orbit angles will be able 
to shed light on their formation mechanisms and dynamical histories. 

\section*{Acknowledgements}
We gratefully acknowledge the \emph{TRAPPIST} team, in particular Micha\"{e}l Gillon, 
and thank them for use of the TRAPPIST-1 planetary system parameters. We also thank 
Scott Gaudi and Joshua Winn for their useful discussions and the anonymous referee 
for their suggestions which we believe greatly improved the manuscript. RC also 
thanks the Canadian Institute for Theoretical Astrophysics for use of the Sunnyvale 
computing cluster throughout this work. RC is supported in part by a Centre for 
Planetary Sciences Graduate Fellowship. 

\bibliographystyle{mnras}
\bibliography{refs}

\begin{table*}
\caption{Adopted Stellar and Planet Parameters for the TRAPPIST-1 Planetary 
System.}
\label{trappist1table}
\begin{tabular}{ccccc}
  \hline
  \textbf{Parameter} & & \textbf{Value} & & \textbf{Reference} \\
  \hline
  &  & \textbf{TRAPPIST-1} & & \\
  \hline
  Stellar mass, $M_s$ & & $0.080 \pm 0.009$ M$_{\odot}$ & & \cite{gillon16} \\
  Stellar radius, $R_s$ & & $0.117 \pm 0.004$ R$_{\odot}$ & & \cite{gillon16} \\
  Project rotation velocity, \vsini{} & & $6 \pm 2$ km s$^{-1}$ & & 
  \cite{reiners10b} \\
  Rotation period, P$_{\mathrm{rot}}$ & & $1.40 \pm 0.05$ days & & \cite{gillon16}\\
  Linear limb-darkening coefficient, $u_1^a$ & & $0.021$ & & \cite{claret11}\\
  & & & & \cite{eastman13}\\
  Quadratic limb-darkening coefficient, $u_2^a$ & & $0.376$ & & \cite{claret11}\\
  & & & & \cite{eastman13} \\
  \hline
  & \textbf{TRAPPIST-1b} &\textbf{TRAPPIST-1c} &\textbf{TRAPPIST-1d} &\\
  Orbital period, $P$ (days) & $1.510848$ & $2.421848$ & $18.20200$ & \cite{gillon16} \\
  Time of mid-transit, $T_0$ (BJD) & 2457322.51765 & 2457362.80520 & 2457294.7744 & 
  \cite{gillon16} \\
  Planetary radius, $r_p$ (R$_{\oplus}$) & 1.113 & 1.049 & 1.163 & \cite{gillon16}\\
  (assumed) Planetary mass, $m_p$ (M$_{\oplus}$) & 1.379 & 1.154 & 1.573 & This work\\
  Eccentricity, $e$ & 0 & 0 & 0 & \cite{gillon16} \\
  Inclination, $i_p$ (deg) & 89.410 & 89.488 & 89.893 & \cite{gillon16}\\
  Impact parameter, $b$ ($R_s$) & 0.21 & 0.25 & 0.24 & \cite{gillon16}\\
  (expected) Doppler semi-amplitude, $K_{\mathrm{Dop}}$ (m s$^{-1}$) & 4.136 & 2.959 & 2.059 & 
  This work\\
  (expected) Rossiter-McLaughlin semi-amplitude, &&&& \\ 
  $K_{\mathrm{RM}}$ (m s$^{-1}$) & 45.909 & 40.747 & 50.162 & This work\\
  \hline
  \multicolumn{5}{p{\linewidth}}{$^a$ Obtained from the interpolation over the grid of quadratic 
    limb-darkening coefficients from \cite{claret11} in the J-band. For the star we 
    assume $T_{\mathrm{eff}} = 2550$ K, [Fe/H]= 0.04 dex, and $\log{g}=5.2$ (cgs) 
    \citep{gillon16}.}
\end{tabular}
\end{table*}

\bsp
\label{lastpage}
\end{document}